\documentclass{article}
\usepackage{epsfig}
\usepackage{amsmath}
\usepackage{amsfonts}\tolerance=10000
\date{\today} 

 \def\D{\Delta}

\def\D{\Delta}

\usepackage{graphicx}
\usepackage{cite}
\usepackage{amssymb}
\usepackage{color}

\newcommand{\insertplot}[5]{\begin{figure}
 \hfill\hbox to 0.05in{\vbox to #5in{\vfill
 \inputplot{#1}{#4}{#5}}\hfill}
 \hfill\vspace{-.1in}
 \caption{#2}\label{#3}
 \end{figure}}
 \newcommand{\inputplot}[3]{
 \special{ps: plotfile #1}
}
\newcounter{fig}

\newcommand{\beq}{\begin{equation}}
\newcommand{\eeq}{\end{equation}}
\newcommand{\beqs}{\begin{eqnarray}}
\newcommand{\eeqs}{\end{eqnarray}}

\numberwithin{equation}{section}
\newcommand{\be}{\begin{equation}}
\newcommand{\ee}{\end{equation}}
\newcommand{\bea}{\begin{eqnarray}}
\newcommand{\eea}{\end{eqnarray}}

\usepackage{graphicx}

\begin{document}
\begin{center}

{\Large \bf 
Two Schwarzschild-like black holes \\ balanced by their scalar hair
} 
\vspace{0.8cm}
\\
{{\bf
Carlos A. R. Herdeiro
 and
Eugen Radu
}
\vspace{0.3cm}
\\
$^{\ddagger }${\small Departamento de Matem\'atica da Universidade de Aveiro and } \\ {\small  Centre for Research and Development  in Mathematics and Applications (CIDMA),} \\ {\small    Campus de Santiago, 3810-183 Aveiro, Portugal}
}
\vspace{0.3cm}
\end{center}

\begin{abstract}
%
We show that, unlike vacuum General Relativity, Einstein-scalar theories allow \textit{balanced} static, neutral, asymptotically flat, double-black hole solutions, for scalar field models minimally coupled to gravity, with appropriate self-interactions. These are scalar hairy versions of the double-Schwarzschild (or Bach-Weyl) solution, \textit{but regular} on and outside the two (topologically spherical) horizons. 
The balancing repulsive force  
 is provided by the scalar field.  An explicit illustration is presented, using a Weyl-type construction adapted to numerical solutions, requiring no partial linearisation, or integrability structure, of the Einstein-scalar equations. 
Fixing the couplings of the model, the balanced configurations
form a one-parameter family of solutions,
labelled by the proper distance 
between the black holes.
\end{abstract}

\tableofcontents

\newpage

\section{ Introduction}
 
A remarkable solution of ``vacuum" General Relativity (GR) is the Bach-Weyl (BW) or double-Schwarzschild metric, 
describing two static, neutral black holes (BHs) 
 placed at some non-zero distance,
in a four dimensional, asymptotically flat spacetime~\cite{BW}.\footnote{The quotes in ``vacuum" emphasise that even if one is solving the vacuum Einstein equations, the existence of conical singularities implies localised sources.}  
The gravitational attraction between the BHs is unbalanced; as a result,
conical singularities along the symmetry axis are mandated by the field equations~\cite{Einstein:1936fp}.
Despite such naked singularities, this solution has a well defined gravitational
action. Moreover,  the Bekenstein-Hawking area law still holds when 
 using standard Euclidean gravity thermodynamical arguments
\cite{Herdeiro:2009vd}.

The precise location of the conical singularity of the BW solution is a matter of choice. 
It  can either be chosen in between the two BHs - in which case it is interpreted as a \textit{strut} - or connecting either BH to infinity - in which case it is interpreted as two \textit{strings}. 
In order for the spacetime to be asymptotically flat, without any conical singularities at spatial infinity, 
one often takes the former viewpoint. 
Then, the strut energy is interpreted 
as the interaction energy between the BHs, while its pressure prevents 
the gravitational collapse of the system
\cite{Costa:2000kf}.

On the one hand, the BW solution can be generalised within ``vacuum" GR in different ways. One way is to introduce $N$ (instead of 2) colinear, neutral, static BHs, leading to the   Israel-Kahn   solution~\cite{IK}. However, this does not solve the need for a conical singularity, except in the $N\rightarrow \infty$ limit~\cite{Myers:1986rx}, which has a natural interpretation as a BH in a compactified spacetime, rather than an asymptotically flat configuration. Another way is to place the BW solution in an appropriate external gravitational field~\cite{Astorino:2021dju,Vigano:2022hrg}. Such solution ceases, again, to be asymptotically flat; in fact it is plagued by naked curvature singularities at spatial infinity.\footnote{In~\cite{Astorino:2021dju} a \textit{local} perspective is taken, to argue on the physical merits of such solutions.} A final way is to make the BHs \textit{spin}. The double-Kerr solution can be constructed via elaborate solution generating techniques, such as the inverse scattering method~\cite{KramerN}. For co-rotating BHs, with aligned spins, the spin-spin interaction is repulsive~\cite{Wald:1972sz}, introducing a plausible  balancing effect. It turns out, however, that this extra interaction cannot balance the system, for objects covered by an event horizon - see $e.g.$~\cite{Hennig:2019knn}. A physical explanation has been put forward in~\cite{Costa:2009wj}.

On the other hand, the BW solution can be generalised within \textit{electrovacuum} GR to yield balanced, asymptotically flat configurations. This involves making the BHs \textit{extremal}, $i.e.$ with their maximal charge to mass ratio. The corresponding balanced BHs fall into the Majumdar-Papapetrou class of metrics~\cite{Majumdar:1947eu,Papapetrou:1948jw}, describing $N$ extremal Reissner-Nordstr\"om BHs in equilibrium~\cite{Hartle:1972ya}, which are regular on and outside the event horizon and asymptotically flat. Such solutions can also be generalized to Einstein-Maxwell dilatonic theories - see $e.g.$~\cite{Gibbons:1987ps,Cornish:1996de,Chen:2012dr}. There are also non-asymptotically flat charged BHs in equilibrium, when immersed in a Melvin-type universe~\cite{Emparan:2001bb} or in a de Sitter Universe~\cite{Kastor:1992nn}; in the latter case the BHs are co-moving with the cosmological expansion.

\medskip

To the best of our knowledge, no static, electro-magnetically neutral, asymptotically flat BHs in equilibrium are known in four spacetime dimensions.\footnote{In higher dimensions, there are vacuum multi-BH solutions, like the black Saturn~\cite{Elvang:2007rd}, allowed by the non-trivial topology of the event horizons permitted by higher dimensional vacuum gravity~\cite{Emparan:2001wn}. These solutions are stationary, rather than static.} There is no reason, however, to expect this to be a fundamental feature of relativistic gravity. Conceptually, the electromagnetic repulsive interaction that allows balance in some of the aforementioned solutions could, in principle, be replaced by another repulsive interaction, namely scalar. Technically, however, one faces important obstacles. 

For solutions akin to the Majumdar-Papapetrou solution (multi-BHs experiencing a ``no-force" condition), common in supergravity theories (see $e.g.$~\cite{Duff:1994an}), under an appropriate ansatz one observes a \textit{full} linearization of the Einstein-matter equations. This allows a superposition principle that corresponds to adding multiple BHs and it is intimately connected with supersymmetry~\cite{Gibbons:1982fy,Tod:1983pm}. For solutions akin to the double-Schwarzschild metric, under an appropriate ansatz - corresponding to the Weyl formalism \cite{Weyl}, one observes a \textit{partial} linearization of the full Einstein equations. This still allows a superposition principle for a specific metric function, which in effect corresponds to adding multiple BHs, with the remaining metric functions obeying non-linear equations, which can, nonetheless, be straighforwardly solved once the linear metric function function is known.  The Weyl formalism comes with an intuitive diagramatic construction - the rod structure (see $e.g.$~\cite{Emparan:2001wk,Harmark:2004rm}) - that permits constructing new solutions. The static Weyl solutions constructed in this way, moreover, serve as natural seeds for the inverse scattering technique~\cite{Belinsky:1971nt}, that can add rotation (and other properties, such as NUT charges) to the solutions.

\medskip

For scalar fields with canonical kinetic terms, minimally coupled to Einstein's theory, possibly with some self-interacting potentials -- hereafter \textit{Einstein-scalar theories} --, the Weyl construction has not been made to work, except in the case of free, massless scalar fields~\cite{Belinsky:1979wi,Astorino:2014mda}. In the absence of a methodology to obtain exact multi-BH solutions, one may approach
 such configurations numerically,
as solutions of partial differential equations (PDEs) with suitable boundary conditions.
This paper aims at proposing a general 
framework for
the  study of static multi-BH systems with matter fields, numerically.
 As an application, we shall report solutions describing two balanced BHs (hereafter dubbed \textit{2BHs}) in a specific Einstein-scalar theory. 

\medskip

The construction we propose is, in principle, more general than scalar matter models. The choice of a scalar field for the matter content is mainly motivated by 
its simplicity, both technical and  conceptual.
Simultaneously, the
influential theorem by Bekenstein 
\cite{Bekenstein:1972ny}
forbidding the existence of (single) BHs with scalar hair
can be circumvented in different ways~\cite{Herdeiro:2015waa}.
A simple way 
is to allow a scalar field potential
which is not strictly positive, such that the  
energy conditions assumed by the theorem are violated \cite{Herdeiro:2015waa}.
Such scalar fields can provide an extra repulsive interaction, balancing a non-trivial scalar field profile  outside the 
horizon.
(Single) BHs with scalar hair are allowed by this mechanism, $e.g.$~\cite{Nucamendi:1995ex,Gubser:2005ih,Kleihaus:2013tba,Bakopoulos:2021dry}.

One may thus anticipate the 
 same mechanism to work for the 2BHs case as well.
A scalar field potential which can take negative values in the 
region between the horizons, moreover,
could provide the extra (repulsive) interaction to balance two neutral static BHs, 
curing the conical singularity of the BW solution. 
Indeed, this is confirmed by the results  in this work, 
where we present
numerical evidence for the existence of balanced 2BHs solutions in this setting. A central point in our approach is that the rod structure of the BW
solution  can be used also for such Einstein-matter configurations,
in particular for the 2BHs system with scalar hair, even though the partial linearization of the Einstein-scalar equations and the vacuum 
Newtonian interpretation of the rods cease to be valid. 
The application given in this work will focus on 2BH systems in thermal equilibrium,
$i.e.$
with  two identical BHs placed at some distance.

\medskip

This paper is organized as follows. In Section~\ref{model} we present the Einstein-scalar model we shall work with. The canonical Weyl construction is attempted with this model, to observe the known obstructions. Then, we specialize to vacuum to discuss the Schwarzschild, double-Schwarzschild (or BW) solutions and the rod structure. In Section~\ref{formalism} we present a Weyl-type construction adapted to numerics, readdressing the rod structure, discussing the boundary conditions and the single BH limit. In Section~\ref{results} we specialize to a potential that allows BHs to have scalar hair and thus that may allow such hair to balance two BHs. 
We then report the results for such 2BHs balanced system. 
Some final remarks close this paper, which also contains 
three appendices with technical details,
together with a numerical construction of the double-Reissner-Nordstr\"om solution (2RNBHs), as a
test of the proposed numerical scheme.

\section{Einstein-scalar models and the vacuum Weyl construction}
\label{model}
\subsection{The action and equations }
 \label{Saction}

We consider the Einstein-scalar model described by the following action
\begin{eqnarray}
\label{action}
 \mathcal{S}=
\frac{1}{4 \pi}
\int d^4 x
\sqrt{-g}
\left[
\frac{R}{4 G}
   -\frac{1}{2} g^{\alpha\beta}
	          \left(
	\Phi_{, \, \alpha}^* \Phi_{, \, \beta} 
	+ \Phi _{, \, \alpha}^* \Phi _{, \, \beta}
            \right)
						          -U(\left|\Phi\right|)
\right]~,
\end{eqnarray}
where $G$ is the gravitational constant, $R$ is the Ricci scalar associated with the
spacetime metric $g_{\alpha\beta}$, which has determinant $g$, 
$\Phi$ is a complex scalar field with $^*$ denoting complex conjugation  and
  $ U(|\Phi|)$ denotes the scalar potential.
The scalar field mass is defined by  $\mu^2\equiv  (d^2 U/d |\Phi|^2)\big|_{\Phi=0}$.
 
The Einstein--scalar field equations, 
obtained by varying (\ref{action}) with respect to the metric and scalar field   are, respectively,
\begin{eqnarray}
\label{eqs}
E_{\alpha\beta} 
\equiv 
R_{\alpha \beta}-\frac{1}{2}g_{\alpha \beta}R-2 G  T_{\alpha \beta}=0 \ , \qquad  
~~~
\nabla_{\alpha}\nabla^{\alpha}\Phi=\frac{d U}{d\left|\Phi\right|}  \ ,
\end{eqnarray}  
where $T_{\alpha\beta}$ is the
 energy-momentum tensor  of the scalar field
\begin{eqnarray}
\label{tmunu} 
T_{\alpha\beta} 
=
\partial_\alpha\Phi^* \partial_\beta\Phi 
+\partial_\beta\Phi^* \partial_\alpha\Phi  
-g_{\alpha\beta}  \left[ \frac{1}{2} g^{\gamma\delta} 
 ( \partial_\gamma \Phi^* \partial_\delta\Phi+
\partial_\delta\Phi^* \partial_\gamma\Phi) +U(\left|\Phi\right|) \right]
 \ .
\end{eqnarray}

\subsection{Canonical Weyl-coordinates}
\label{ansatz}

The configurations we shall consider herein are static and axially symmetric,  admitting two
orthogonal, commuting, non-null Killing Vector Fields (KVFs).
In what follows we take their line element written in coordinates adapted these symmetries, $(t,\rho,z,\varphi)$, such that $\partial_t$ and $\partial_\varphi$ are KVFs; it reads
\begin{eqnarray}
\label{metric}
ds^2=- e^{2\mathcal{U}(\rho,z)}dt^2+e^{-2\mathcal{U}(\rho,z)}\left[e^{2\mathcal{K}(\rho,z)} (d\rho^2+dz^2)+e^{2\mathcal{C}(\rho,z)}\rho^2d\varphi^2\right] \  ,
\end{eqnarray}
thus introducing three unknown functions 
$\mathcal{U}$, $\mathcal{K}$ and $\mathcal{C}$
 of the non-Killing coordinates $(\rho,z)$, 
where   $0\leqslant \rho<\infty,$ $-\infty< z<\infty$
and $0\leqslant \varphi  < 2\pi $. 
For the scalar field $\Phi$, we take a generic ansatz
\begin{eqnarray}
\label{scalar}
 \Phi=\phi(\rho,z) e^{i m \varphi}\ ,
\end{eqnarray}
with $\phi$ a real function --  the scalar field amplitude --,
and $m\in \mathbb{Z}$.

Appropriate combinations 
of the Einstein equations,
 $E_t^t=0$,
$E_\rho^\rho+E_z^z=0$
and
$E_{\varphi}^{\varphi}=0$,
yield the following set of equations for the functions 
$\mathcal{U}$,
$\mathcal{K}$, and
$\mathcal{C}$, 
 \begin{eqnarray}
\nonumber
&&
\Delta \mathcal{U}  
+(\nabla  \mathcal{U})\cdot( \nabla  \mathcal{C})
=- 2 G e^{2( \mathcal{K}- \mathcal{U})} U(\phi) \ ,
\\
\nonumber
&&
\Delta \mathcal{K} 
-\frac{\mathcal{K}_{,\rho}}{\rho} 
+(\nabla \mathcal{U}  )^2 
=-2G 
\left[
(\nabla \phi)^2
+e^{2( \mathcal{K}- \mathcal{U})} U(\phi)
-\frac{e^{2( \mathcal{K}- \mathcal{C})}m^2\phi^2}{\rho^2}
\right] \ ,
\\
\label{eqs1s} 
&& 
\Delta \mathcal{C} 
+\frac{\mathcal{C}_{,\rho}}{\rho} 
+(\nabla \mathcal{C}  )^2 = - 4 G e^{2( \mathcal{K}- \mathcal{U})} 
\left[
U(\phi)
+\frac{e^{2( \mathcal{U}- \mathcal{C})}m^2\phi^2}{\rho^2}
\right]
\ .
\end{eqnarray}
The equation for the scalar field amplitude $\phi$
is
 \begin{eqnarray}
\label{eqs2} 
\Delta \phi
+ (\nabla \mathcal{C})\cdot( \nabla \phi) 
=\frac{1}{2}e^{2( \mathcal{K}- \mathcal{U})} \frac{dU(\phi)}{d\phi}
+\frac{e^{2( \mathcal{K}- \mathcal{C})}m^2\phi }{\rho^2}
\ ,
 \end{eqnarray}
We have defined, acting on arbitrary functions $\mathcal{F}(\rho,z)$ and  $\mathcal{G}(\rho,z)$, 
\begin{eqnarray}
\label{rel}
(\nabla \mathcal{F}) \cdot (\nabla \mathcal{G}) \equiv  \mathcal{F}_{,\rho}  \mathcal{G}_{,\rho}+  \mathcal{F}_{,z}  \mathcal{G}_{,z} \ , \qquad 
\Delta \mathcal{F}\equiv \mathcal{F}_{,\rho\rho}+ \mathcal{F}_{,z z} +\frac{1}{\rho} \mathcal{F}_{,\rho}\ .
\end{eqnarray}
These operators are the covariant operators on an auxiliary Euclidean 3-space in standard cylindrical coordinates, $ds^2_{\rm auxiliary}=d\rho^2+\rho^2d\varphi^2+dz^2$.

The remaining Einstein equations 
$E_\rho^z=0,~E_\rho^\rho-E_z^z=0$
yield two constraints,
\begin{eqnarray}  
\nonumber
&&
\mathcal{C}_{,\rho \rho}
+\frac{2\mathcal{C}_{,\rho }}{\rho}
-\mathcal{C}_{,zz}
+2(\mathcal{U}_{,\rho}^2-\mathcal{U}_{,z}^2)
+ \mathcal{C}_{,\rho}^2-\mathcal{C}_{,z}^2
-2(\mathcal{C}_{,\rho}\mathcal{K}_{,\rho}-\mathcal{C}_{,z}\mathcal{K}_{,z} )
-\frac{2 \mathcal{K}_{,\rho}}{\rho}
+4 G (\phi_{,\rho}^2-\phi_{,z}^2)=0 \ ,
\\
\label{constraints}
&&
 \mathcal{C}_{,\rho z}
+ \mathcal{C}_{,\rho} \mathcal{C}_{,z}
+2  \mathcal{U}_{,\rho} \mathcal{U}_{,z}
-\mathcal{C}_{,\rho} \mathcal{K}_{,z}
-\mathcal{K}_{,\rho} \mathcal{C}_{,z}
+\frac{\mathcal{C}_{,z} }{\rho}
-\frac{\mathcal{K}_{,z} }{\rho}
+4 G \phi_{,\rho}\phi_{,z}=0 \ .
\end{eqnarray}
 Following \cite{Wiseman:2002zc}, we note that
setting $E^t_t=E^{\varphi}_{\varphi} =E^\rho_\rho+E^z_z=0$
in $\nabla_\alpha E^{\alpha \rho}=0$ and $\nabla_\alpha E^{\alpha z}=0$,  
we obtain a set of  Cauchy-Riemann relations.
 Thus the weighted constraints satisfy Laplace equations,
and the constraints are fulfilled,
when one of them is satisfied on the boundary 
and the other at a single point
\cite{Wiseman:2002zc}.

\subsection{Scalar-vacuum}
\label{scalarvacuum}

In the absence of a scalar potential, $i.e.$ $U(\phi)=0$,  
and for a real scalar ($m=0$),
the third equation in~\eqref{eqs1s} allows us to take $\mathcal{C}=0$. The problem then has only three unknown functions, $\mathcal{U}$,
$\mathcal{K}$, and
$\phi$. Two of them obey linear equations. Indeed,  
the first equation in~\eqref{eqs1s} and the scalar equation become Laplace-type equations
\begin{equation}
\Delta \mathcal{U}=0 \ , \qquad \Delta \phi=0 \ .
\label{laplace}
\end{equation}
 This is the aforementioned partial linearization of the Einstein equations. It is easy to see from~\eqref{eqs1s} that this linearization is lost 
for $m\neq 0$
or
in the presence of a potential $U(\phi)$, 
which leads to a nuisance (from the viewpoint of the Weyl construction) source term. 
The source term is, however, absent for a  free, massless real scalar. 
The Weyl construction has in fact been explored in that case~\cite{Belinsky:1979wi,Astorino:2014mda}. 

Instead of determining the remaining function $\mathcal{K}(\rho,z)$ from the second eq. in~\eqref{eqs1s}, it can be determined from the two constraint equations~\eqref{constraints}, which reduce to 
\begin{eqnarray}  
\mathcal{K}_{,\rho}=\rho (\mathcal{U}_{,\rho}^2-\mathcal{U}_{,z}^2)
+4 G\rho (\phi_{,\rho}^2-\phi_{,z}^2) \ , \qquad 
\label{constraints2}
&&
\mathcal{K}_{,z} =2\rho  \mathcal{U}_{,\rho} \mathcal{U}_{,z}
+4 G\rho \phi_{,\rho}\phi_{,z} \ .
\label{lineint}
\end{eqnarray}
Thus, once $\mathcal{U},\phi$ are determined by solving the linear equations~\eqref{laplace}, then $\mathcal{K}$ is determined by solving two line integrals from~\eqref{lineint}. With what concerns BHs, this scalar-vacuum Weyl construction has not been rewarding, as in scalar vacuum BH hair is forbidden~\cite{Bekenstein:1972ny,Herdeiro:2015waa}. As such, let us focus on the particular case of vacuum ($\phi=0$).

\subsection{Vacuum: Schwarzschild and rod-structure}
\label{vacuum}

Setting $\phi=0$ in~\eqref{laplace} and~\eqref{lineint}, the simplest solution has $\mathcal{U}=0=\mathcal{K}$, which is Minkowski spacetime. We are now interested in asymptotically Minkowski solutions.

The Schwarzschild BH with mass $M$ appears in this Weyl construction as the Newtonian potential\footnote{The Laplace eq. $\Delta \mathcal{U}=0$ has no source, but it can be regarded as the Poisson equation of Newtonian gravity with sources along the $z$ axis only, wherein the Laplace operator in cylindrical coordinates is not defined.} $\mathcal{U}$ of an infinitely thin rod of length $2M$ along the $z$-axis. Placing this rod symmetrically w.r.t. $z=0$ between $z=-M$ and $z=M$, this means that
\begin{equation}
e^{2\mathcal{U}}=\frac{r_++r_--2M}{r_++r_-+2M} \ , \qquad r_\pm=\sqrt{\rho^2+\left(z\pm M\right)^2} \ .
\label{schwey1}
\end{equation}
Additionally, from~\eqref{lineint}
\begin{equation}
e^{2\mathcal{K}}=\frac{(r_++r_-)^2-(2M)^2}{4r_+r_-}  \ .
\label{schwey2}
\end{equation}
The standard Schwarzschild coordinates $(t,r,\theta,\varphi)$ can be recovered from the transformation $(\rho,z)\rightarrow (r,\theta)$ via
\begin{equation}
\rho=\sqrt{r^2-2Mr}\sin\theta \ , \qquad z=(r-M)\cos\theta \ .
\label{ct1}
\end{equation}

 A large class of solutions in Weyl coordinates are characterized
by the boundary conditions on the $z-$axis, 
known as the {\it rod-structure}.
That is, the $z-$axis is divided into 
$N$ intervals 
(called rods of the 
solution), $[-\infty, z_1]$,
$[z_1,z_2]$,$\dots$, $[z_{N-1},\infty]$.
A necessary condition for a regular solution is that only one of the functions
$g_{tt}(0,z)$
or 
$g_{\varphi\varphi}(0,z)$  
becomes zero for a given rod 
(except for isolated points between the intervals).

\begin{figure}[h!]
\hbox to\linewidth{\hss%
\resizebox{15cm}{4cm}
{\includegraphics{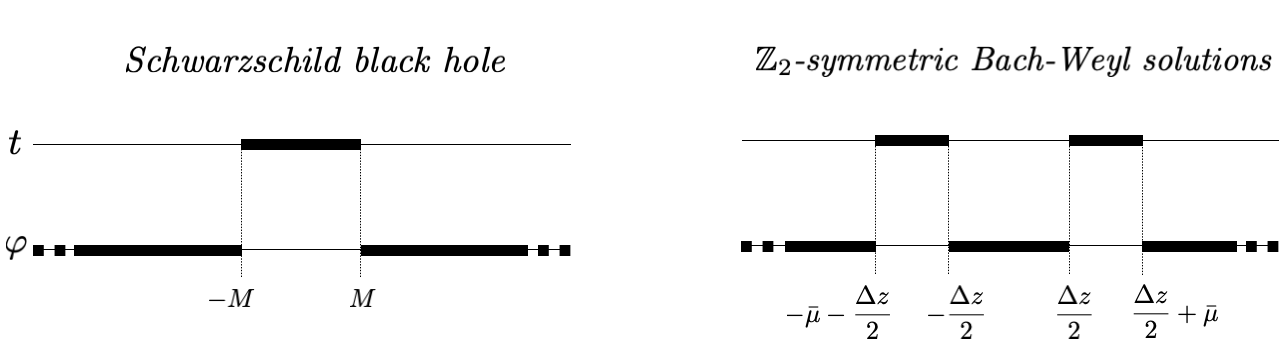}}
\hss}\caption{   
Rod structure, encoding the boundary conditions at $\rho=0$ along the $z$ axis, for the Schwarzschild solution (left) and $\mathbb{Z}_2$-symmetric BW solution (right).
 }
\label{figure1}
\end{figure}

The rods are timelike or spacelike:
\begin{itemize}
\item[$\bullet$] Event horizons are described by timelike rods. They  are sets of fixed points of the $\partial_t$ KVF with $g_{tt}(0,z)=0$ and
\begin{equation}
\lim_{\rho\to 0}\frac{g_{tt}(\rho,z)}{\rho^2}<0 \ .
\end{equation}
\item[$\bullet$] The symmetry axes are described by spacelike rods. They are sets of fixed points of the $\partial_\varphi$ KVK, with  $g_{\varphi\varphi}(0,z)=0$ and
\begin{equation}
\lim_{\rho\to 0}\frac{g_{\varphi\varphi}(\rho,z)}{\rho^2}>0 \ .
\end{equation}
\end{itemize}
Fig.~\ref{figure1} (left panel) exhibits the rod structure of the Schwarzschild solution.

\subsection{Vacuum: $\mathbb{Z}_2$-symmetric Bach-Weyl}
\label{vacuum2}

The rod structure allows an intuitive, diagramatic-based reconstruction of the metric. For instance, one can easily imagine the rod structure of the double-Schwarzschild (or BW) solution - Fig.~\ref{figure1} (right panel). Choosing the location of the horizons $\mathbb{Z}_2$-symmetric with respect to $z=0$, both  at $\rho=0$, with the ``lower" one at $-\Delta z/2 -\bar  \mu\leqslant z\leqslant -\Delta z/2$
and the ``upper" one at
$ \Delta z/2 \leqslant z\leqslant  \Delta z/2+\bar  \mu$, the $g_{tt}$ metric function is defined by the corresponding Newtonian potential, which reads:
\begin{equation}
\label{BW1}
e^{2\mathcal{U}}=
\frac{(r_1+r_2- \bar \mu)}{(r_1+r_2+ \bar  \mu)}\frac{(r_3+r_4- \bar  \mu)}{(r_3+r_4+ \bar  \mu)} \ ,  
\end{equation}
where 
\begin{eqnarray}
\label{Ri}
&&
r_1=\sqrt{\rho^2+\left(z-\frac{\Delta z}{2}- \bar  \mu\right)^2},~~ 
r_2=\sqrt{\rho^2+\left(z-\frac{\Delta z}{2}\right)^2},~~ 
\\
\nonumber
&&
r_3=\sqrt{\rho^2+\left(z+\frac{\Delta z}{2} \right)^2},~~ 
r_4=\sqrt{\rho^2+\left(z+\frac{\Delta z}{2} +\bar  \mu\right)^2}.
 \end{eqnarray}
Again, from~\eqref{lineint}
\begin{equation}
e^{2\mathcal{K}}=
\left(\frac{\Delta z}{\Delta z+ \bar  \mu}\right)^2
\left(\frac{(r_1+r_2)^2-\bar  \mu^2}{4r_1r_2}\right)
\left(\frac{(r_3+r_4)^2-\bar  \mu^2}{4r_3r_4}\right)
\left(\frac{(\Delta z+ \bar  \mu) r_1+(\Delta z
+2 \bar  \mu)r_2-\bar  \mu r_4}{\Delta z~r_1+(\Delta z+ \bar \mu)r_2- \bar  \mu r_3}\right)^2 \ .
\label{BW2}
\end{equation}

This 2-parameter solution\footnote{We follow the conventions in~\cite{Costa:2000kf}.} describes two equal BHs 
 in thermodynamical equilibrium - with the same Hawking temperature and horizon area.
The parameter $\bar \mu$ is
the ADM mass of this spacetime (twice the individual BH masses):
\begin{eqnarray}
M=\bar  \mu>0 \ .
 \end{eqnarray}
The parameter 
 $\Delta z \geqslant 0$ 
provides
the coordinate distance between the two horizons along the $z$ axis.

This system has a deficit angle along the section in between the BHs,
$i.e.$ for $-\Delta z/2 \leqslant z\leqslant \Delta z/2$,
with a strength 
$\delta$, 
as defined by the relation (\ref{delta}) below, given by:
\begin{eqnarray}
\frac{\delta}{2\pi}=-\frac{\bar  \mu^2}{(\D z+2\bar  \mu)\D z}<0 \ .
\end{eqnarray}
The proper distance
between the BHs is
\begin{eqnarray}
L=\int_{-\Delta z/2}^{\-\Delta z/2}dz f_1(0,z) =
\Delta z \left(\frac{u+4}{u+2}\right)^2 E(\bar m) \ ,
~~~{\rm where}~~
u=\frac{2\Delta z}{\bar \mu} \ ,~~\bar m=\left(\frac{u }{ u+4}\right)^2 \ , 
\end{eqnarray}
$E(\bar m)$ being the complete elliptic integral of the second kind.
The event horizon area of each BH and the corresponding Hawking temperature are:
\begin{eqnarray}
A_H= 4 \pi \bar  \mu^2 \frac{\Delta z+2 \bar \mu }{\Delta z+\bar \mu} \ , \qquad 
T_H=\frac{1}{4 \pi \bar  \mu}\frac{\Delta z+\bar  \mu}{\Delta z+2 \bar  \mu} \ .
\end{eqnarray}
%
%
For $\Delta z=0$ the two BH horizons coalesce,
and we are left the (single) Schwarzschild BH in Weyl coordinates.
Also, the solution trivializes as $\bar \mu \to 0$,
a limit which corresponds to flat spacetime.

We remark that the solution above 
captures already all the basic  features of its Israel-Kahn
generalization \cite{IK},
with $N$ horizons placed arbitrarily 
on a common symmetry axis.
However,  the $N>2$ metric functions
get increasingly more involved, albeit with an underlying common structure.

\section{A Weyl-type construction adapted to numerics} 
\label{formalism}

\subsection{The  rod structure and quantities of interest }
 \label{generic}

The Weyl construction does not carry through to generic matter models, as illustrated in the previous section for the Einstein-scalar models with a potential.  In this work, however, we argue that the Weyl coordinates in (\ref{metric}) 
together with the rod
structure 
of the vacuum multi-BH solutions
can be used to construct physically relevant solutions beyond the simplest theories where the Weyl construction allows a partial linearization of the field equations.
\textit{A priori}, this
is not guaranteed, and the validity of
the metric ansatz could be proven only \textit{a posteriori}, after solving the field equations.
Of course, the elegant Newtonian interpretation of the rods is lost in the non-(scalar, electro)vacuum case; but a working method allows exploring the physics of the non-linear solutions.

Even though one could hold on to the canonical metric parameterization in the Weyl ansatz~(\ref{metric}), a numerical implementation, namely the boundary conditions at the horizons, is facilitated by taking the simpler parameterization of the metric functions
\begin{eqnarray}
\label{metric2} 
ds^2=- f_0(\rho,z)dt^2+f_1(\rho,z)(d\rho^2+dz^2)+f_2(\rho,z)d\varphi^2  \ ;
\end{eqnarray}
in other words, we relabel 
\begin{equation}
f_0(\rho,z)\equiv e^{2\mathcal{U}(\rho,z)} \ , \qquad  f_1(\rho,z)\equiv e^{-2\mathcal{U}(\rho,z)+2\mathcal{K}(\rho,z)} \ , \qquad f_2(\rho,z)\equiv \rho^2e^{-2\mathcal{U}(\rho,z)+2\mathcal{C}(\rho,z)} \ .
\label{dictionary}
\end{equation}
 For completeness, the field equations in terms of this parameterization are given in Appendix~\ref{appendix1}. With this parameterization
the following expressions of the metric functions 
and scalar field 
near the $z-$axis
are compatible with the Einstein-scalar field
equations:
\begin{eqnarray}
\label{rods}
 f_i(\rho,z)=f_{i0}(z)+\rho^2f_{i2}(z)+\dots \ , \qquad 
\phi(\rho,z)=\phi_{0}(z)+\rho^2 \phi_{2}(z)+\dots \ ,
\end{eqnarray}
where the functions 
$f_{i0}(z)$, 
$f_{i2}(z)$ 
 (with $i=0,1,2$)
and
$\phi_{0}(z)$,
$\phi_{2}(z)$
satisfy a complicated set 
of nonlinear second order ordinary differential equations.

Our main assumption 
(supported by the results reported below)
is that,
similarly to the vacuum case,
the $z-$axis is divided into $N$ intervals: the rods of the solution.
Moreover, we assume that,
except for isolated points between the rods,   
only one of the functions
$f_0(0,z)$
or 
$f_2(0,z)$  becomes zero for a given rod, while 
the remaining functions stay finite at $\rho = 0$, in general.
Also, one imposes the condition that the 
union of the 
$N$ intervals  covers the entire $z$-range. There are again timelike and spacelike rods, which we now discuss separately.

A finite timelike rod corresponds to an event horizon,
which we assume is
located for  $z_{H_1} \leqslant z\leqslant z_{H_2}$.
Therein, one can further specify the generic expansion~\eqref{rods} to have $f_{00}(z)=0$, such that\footnote{For several horizons,  one should write
such an expansion for each of them.}
\begin{eqnarray}
f_0(\rho,z)=\rho^2f_{02}(z)+\rho^4f_{04}(z)+\dots \ , 
\label{nrod2}
\end{eqnarray}
 with $\lim_{\rho\to 0}\rho^2 f_1/f_0=$const., 
as implied by the constraint equation $E_\rho^z=0$.
The
 horizon metric is given by
\begin{eqnarray}
\label{hm}
d\sigma^2_H=f_1(0,z)dz^2+f_2(0,z)d\varphi^2 \ .
\end{eqnarray}
Two quantities associated with an event horizon are
the event horizon area $A_H$ and
the Hawking temperature $T_H$; they read
 \begin{eqnarray}
\label{AHTH}
A_H = 2\pi \int_{ z_{H_1} }^{ z_{H_2} }dz \sqrt{f_1(0,z)f_2(0,z) }\ , \qquad 
T_H=\frac{1}{2\pi}\lim_{\rho\to 0} \sqrt{\frac{f_{0}(\rho,z)}{\rho^2 f_{1}(\rho,z)}} \ .
\end{eqnarray}
The horizon has a spherical topology (despite the possible presence of 
conical singularities).
A suggestive way to graphically represent its shape -- which is
generically very different from a round 2-sphere --
is to define an {\it effective}
horizon radius ${\rm R}$  
 \cite{Costa:2000kf},
by introducing an angular variable
$z=z(\theta)$,
such that the horizon metric (\ref{hm})
becomes
 \begin{eqnarray} 
\label{R}
d\sigma^2={\rm R}^2 (\theta) (d\theta^2+\sin^2 \theta d\varphi^2) \ , \qquad {\rm with}~~
{\rm R}=
  \frac{\sqrt{f_2(0,z)}}{\sin \theta} \ ,
\end{eqnarray}
where 
  \begin{eqnarray} 
\theta(z)=2 \arctan \left[ C {\rm exp}\left( {\int_{z_{H1}}^z dx \sqrt{\frac{f_1(0,x)}{f_2(0,x)}}}\right)
\right] \ .
\end{eqnarray}
As with the vacuum BW solution \cite{Costa:2000kf},
the constant $C$ is fixed by by requiring
the horizon to be 
 regular at the pole opposite to the other hole.
Alternatively, one can use the standard approach
developed by Smarr for the Kerr BH \cite{Smarr:1973zz},
and consider an 
isometric embedding of the horizon geometry (\ref{hm}) in $\mathbb{E}^3$.

\medskip

Let us now consider the case of a generic spacelike $\varphi-$rod,
 for $z_{S_1}\leqslant z\leqslant z_{S_2}$.  Therein, one can further specify the generic expansion~\eqref{rods} to have $f_{20}(z)=0$, such that, as $\rho\to 0$:
\begin{eqnarray}
f_2(\rho,z)= \rho^2f_{22}(z)+\rho^4f_{24}(z)+\dots \ . 
\end{eqnarray}

One important feature here is that
the constraint equation $E_\rho^z=0$ implies 
the condition
 $f_{10}(z)/f_{22}(z)=$const., 
$i.e.$
a well-defined periodicity for the  coordinate $\varphi$, albeit not necessarily of $2\pi$.
A periodicity 
different from $2\pi$ 
leads to 
the occurrence of a   
conical singularity. Its strength can be measured
 by means of the quantity
\begin{eqnarray}
\label{delta}
\delta
=
2\pi\left(1-\lim_{\rho\rightarrow 0} \sqrt{\frac{f_{2}(\rho,z)}{\rho^2f_{1}(\rho,z)}}\right) \ .
\end{eqnarray}
Then $\delta>0$ corresponds to a conical deficit,
while $\delta<0$ corresponds to a conical excess.   
As with the ``vacuum" case,
a conical deficit  can be interpreted as a string stretched along 
 a certain segment of the $z-$axis, while a conical  
excess is a strut pushing  apart the rods connected to that segment.
A  rescaling of
$\varphi$  can be used to eliminate possible conical singularities on a
given $\varphi$-rod; but in the generic case, once this is fixed, there remain
conical singularities along other $\varphi$-rods.  
 Since we are interested in asymptotically flat solutions, 
we impose $\delta=0$
for the semi-infinite spacelike rods.
 
Another quantity of interest is the proper
length of a $\varphi$-rod
 \begin{eqnarray}
\label{L}
 L=\int_{z_{S_1}}^{z_{S_2}}dz\sqrt{f_1(0,z)} \ ;
\end{eqnarray}
 for a finite rod, $L$ differs from the coordinate distance $z_{S_1}-z_{S_2}$.

\medskip

Let us now consider global quantities.  For large $(\rho,|z|)$,  the functions $f_i$ should approach the Minkowski background
functions, 
while the scalar field vanishes.
The ADM mass $M$ of the 
solutions can be read off from the asymptotic expression of the metric component $g_{tt}$  
\begin{eqnarray}
\label{gtt}
-g_{tt}=f_0\sim 1-\frac{2 G M}{ \sqrt{\rho^2+z^2}}+\dots~.
\end{eqnarray}

The
balanced
solutions with $N$ horizons satisfy the Smarr relation
\cite{Bardeen:1973gs}
\begin{eqnarray}
\label{Smarr}
M=\frac{1}{2G}T_H \sum_{i=1}^N A_H^{(i)} +M_{(\Phi)}\ , 
\end{eqnarray}
 where
\begin{eqnarray}
\label{Mpsi}
M_{(\Phi)} =- \int d^3 x \sqrt{-g}(2T_t^t-T_\alpha^\alpha) \ , 
\end{eqnarray} 
is the contribution to the
total mass of the matter outside the event horizon. They also satisfy  
the $1^{st}$ law of thermodynamics
 \begin{eqnarray}
 \label{fl}
dM=T_H \frac{1}{4G} \sum_{i=1}^N dA_H^{(i)} \ .
\end{eqnarray} 
%

To measure the hairiness of a configuration we define the 
parameter \cite{Delgado:2016zxv}
 \begin{eqnarray}
 \label{p}
p\equiv \frac{M_{(\Phi)} }{M}~,
\end{eqnarray} 
with $p=0$ in  the vacuum case and
$p=1$ for horizonless configurations.

Finally, let us mention that, as discussed in Section \ref{1BH},
	the single BH limit of this framework 
	leads to results similar to those
	found by employing
	a metric ansatz in term of the
	usual spherical coordinates.

\subsection{The boundary conditions for the 2BHs construction}
\label{2BH}

The above considerations 
allow for a consistent construction of 
 Einstein-scalar field generalizations of the BW solution by solving 
numerically
the field equations (\ref{eqs1}), (\ref{eqs2n}) 
within a non-perturbative approach. 
The presence of an arbitrary number of horizons is automatically imposed by the rod structure, 
leading to a standard boundary value problem.

We assume the rod structure of a generic 2BH system to mimic that of the BW solution considered in Fig.~\ref{figure1}:
a semi-infinite spacelike rod $[-\infty,z_1]$ in the $\varphi$-direction
(with $f_2(0,z)=0$); a first (finite) timelike rod in the interval $[z_1, z_2]$ (with $f_0(0,z)=0$); 
another spacelike rod $[z_2,z_3]$  (with $f_2(0,z)=0$);
a second (finite) timelike rod $[z_3, z_4]$ (with $f_0(0,z)=0$);
finally,
a second semi-infinite spacelike rod along $[z_4,\infty]$ (with $f_2(0,z)=0$),
again in the $\varphi$-direction.
The $\mathbb{Z}_2$-symmetric BW solution
has, in accordance to Fig.~\ref{figure1} (right panel):
 \begin{equation}
z_1=-\Delta z/2 -\bar{\mu} \ , \qquad 
z_2=-\Delta z/2 \ , \qquad
z_3=\Delta z/2 \ , \qquad 
z_4=\Delta z/2+\bar{\mu} \ .
\end{equation}

In practice, we 
have found it convenient to take
\begin{eqnarray}
\label{ans1}
f_i=f_i^{(0)}e^{2F_i} \ ,
\end{eqnarray}
where $f_i^{(0)}$ are background functions, given by the metric functions of the BW solution
 (\ref{BW1})-(\ref{BW2}), with the dictionary~\eqref{dictionary}, 
while $F_i$
are unknown functions encoding the 
corrections to the BW metric. 
The equations satisfied by the $F_i$ can easily be derived from 
(\ref{eqs1})
and we shall not display them here.

 In this approach,
  the functions
 $f_i$
automatically satisfy the
desired rod structure, which are enforced by the use of 
background functions $f_i^{(0)}$,
`absorbing' also the  divergencies associated with coordinate singularities and 
(for $f_2$)
coming from the imposed asymptotic behaviour.\footnote{A qualitatively similar approach has been
used in Ref. \cite{Kleihaus:2009wh}
to construct generalizations of the Emparan-Reall black ring solution
\cite{Emparan:2001wn}.}
We assume that $F_i$ are finite everywhere.

The    boundary conditions satisfied by the metric functions
$F_i$
 are
\begin{eqnarray}
\label{bcn1} 
 \partial_\rho F_i|_{\rho=0}=0 \ , \qquad {\rm for~~}-\infty<z<\infty \ , \qquad 
{\rm and} \qquad F_i=0 \qquad {\rm for} \qquad \rho\to \infty ~~{\rm or }~~z\to \pm \infty \ .
 \end{eqnarray}
Asymptotic flatness imposes
$F_1=F_2$ for the  semi-infinite spacelike rods, while  
$F_1-F_2$ takes a  constant value for a finite spacelike rod.
Moreover, $F_1-F_0$ is constant for a  timelike rod.
The boundary conditions for  the scalar field are   
\begin{eqnarray}
\label{bc-psi}
\nonumber
 \partial_\rho \phi|_{\rho=0}=0\ , \qquad {\rm for~~}-\infty<z<\infty \ , \qquad 
{\rm and} \qquad \phi=0 \qquad {\rm for} \qquad 
\rho\to \infty \qquad {\rm or } \qquad z\to \pm \infty ~,
 \end{eqnarray} 
except for $m\neq 0$ 
(with $m$ the integer in the scalar ansatz (\ref{scalar})),
in which case  one imposes 
\begin{eqnarray}
\label{bc-nn} 
  \phi|_{\rho=0}=0 \ ,~~~~{\rm for~a~} \varphi-{\rm rod} \ .
 \end{eqnarray}

We focus on solutions possessing a $\mathbb{Z}_2$-symmetry,
$i.e.$ with two identical BHs, such that the thermal equilibrium
is guaranteed.
Then the auxiliary metric functions $F_i$
satisfy the condition
$F_i(\rho, -z)$=$F_i(\rho,  z)$,
with Neumann boundary conditions at $z=0$.
The situation with the scalar field is different. 
Although we have found evidence for the existence of
 2BH solutions with an $even$ parity scalar 
field amplitude -- $i.e.$ $ \phi(\rho, -z)=\phi(\rho,  z)$ -- 
all such configuration studied so far 
still possess a conical singularity, $\delta \neq  0$.
On the other hand,  balanced solutions exist for odd-parity scalar fields,
$\phi(\rho, -z)=-\phi(\rho,  z)$,
this being the case for all solutions reported in this work.\footnote{The same model containts single BH solutions with an odd-parity scalar field
and $m\geqslant 0$.
The phase diagram is complicated, 
 and will be reported elsewhere.  
}
The
energy-momentum tensor of the 
scalar field is still
invariant under the transformation
$ z\to -z $,
with the existence of two regions (on the semi-infinite spacelike rods),
where the scalar energy has the strongest support.
The existence of configurations with $\delta =0$
can presumably be attributed to the extra-interaction between these
two distinct constituents.

We have solved the resulting set of four coupled
non-linear elliptic PDEs numerically, subject to the above boundary
conditions.
 Details on the used  numerical methods 
and on a new coordinate system better
suited for the numerical study 
are presented in Appendix \ref{appendix}.

\subsection{The single (spherical) BH limit  in Weyl-type coordinates}
\label{1BH} 

Before addressing the construction of 
2BH solutions,
it is interesting to consider the limit 
of the proposed formalism 
with
 \begin{eqnarray}
\Delta z=0 \ ,
\end{eqnarray} 
$i.e.$
a single BH horizon.
This study is technically simpler,
although it contains already some
basic  ingredients of the general 2BHs case.

For a generic matter content, 
the spherically symmetric solutions
are usually studied 
in Schwarzschild-like coordinates.\footnote{The considerations
in this subsection can easily be generalized for a different metric gauge
choice in (\ref{Schw-like}).} 
A common parameterization is 
\begin{eqnarray}
\label{Schw-like}
ds^2=-N(r)e^{-2\delta(r)}dt^2+\frac{dr^2}{N(r)}+r^2
(d\theta^2+\sin^2 \theta d\varphi^2) \ ,
\end{eqnarray}
where $r$ is a radial coordinate and $0\leqslant \theta \leqslant \pi$.
The event horizon is located at some $r=r_h>0$,
where $N(r_h)=0$ 
and $\delta(r_h)$ is finite.
The metric functions 
$N(r)$
and
$\delta(r)$ 
are found by solving the Einstein-matter field equations.

Any specific geometry written in the form~\eqref{Schw-like} can, however, be transformed into Weyl-like coordinates.
In principle, 
the coordinate transformation between
$(\rho,z)$
in the Weyl-like line element (\ref{metric2})
and
$(r,\theta)$
in (\ref{Schw-like})
is simple enough,
with
\begin{eqnarray}
\label{coord-transf}
\rho= c_0 \sinh T(r) \sin\theta \ , \qquad
z=c_0 \cosh T(r) \cos\theta \ , \qquad
{\rm where} \qquad T(r)\equiv \int \frac{dr}{r \sqrt{N(r) }} \ .
\end{eqnarray}
The constant $c_0$ is usually fixed by imposing that
asymptotically
$\sqrt{\rho^2+z^2}\to r$.
Then, the (generic) 
expressions 
of the metric functions
in 
(\ref{metric2})
read 
\begin{eqnarray}
\label{exprs}
f_0(\rho,z)=N(r)e^{-2\delta(r)} \ ,\qquad
f_1(\rho,z)=\frac{ r^2}{ c_0^2[\cosh^2 T(r)-\cos^2 \theta] }\ , \qquad 
f_2(\rho,z)=r^2 \sin^2 \theta \ , 
\end{eqnarray}
with $r,\theta$ functions of $\rho$ and $z$,
as found from
(\ref{coord-transf}).

For the (simplest) case
of a Schwarzschild solution, in the gauge~(\ref{Schw-like}) it has 
$N(r)=1-2 M/r$
and
$\delta(r)=0$; then one finds (taking $c_0=M$)
\begin{eqnarray}
T(r)=2\log \left( \sqrt{\frac{r}{2 \bar\mu}}+\sqrt{\frac{r}{2 \bar\mu}-1} \right) \ ,
\end{eqnarray}
and the coordinate transformation~\eqref{ct1}. Then, from (\ref{exprs}), one finds, upon using the dictionary~\eqref{dictionary} with $\mathcal{C}=0$, the forms~\eqref{schwey1} and~\eqref{schwey2}, 
%
which is the $\Delta z=0$
limit of the BW solution.

It turns out, however, 
that the 
integral
(\ref{coord-transf})
 which determines
$T(r)$, 
can only 
be 
computed analytically for 
very special cases.
In general, one can find an analytic expression for
$T(r)$ -- and  thus for the coordinate transformation -- 
only
for $r \to r_h$, or for large $r$. 
Assuming that the horizon is non-extremal,
the generic behavior 
 as $r \to r_h$
of the functions $N(r)$
and $\delta(r)$ 
is
\begin{eqnarray}
\label{nh1}
N(r)=N_1 (r-r_h) +{\cal O}(r-r_h)^2 \ , \qquad 
\delta(r)=\delta_0+ {\cal O}(r-r_h) \ ,
\end{eqnarray}
with 
$N_1>0$
and
$\delta_0$ 
model-specific parameters.
Then, from (\ref{coord-transf}),
one finds the following general expressions
 \begin{eqnarray}
\label{supeq1}
T(r)=\frac{2\sqrt{r-r_h}}{r_h \sqrt{N_1}}+\dots \ , \qquad {\rm and} \qquad 
\rho=c_0 \frac{2\sqrt{r-r_h}}{r_h\sqrt{N_1}} \sin \theta \ , \qquad 
z=c_0 \left( 
1+\frac{2(r-r_h)}{r_h^2 N_1}
\right)  \cos \theta \ ,
\end{eqnarray}
which is the leading order result in 
a $(r-r_h)$-expansion.
The same approximation implies
 \begin{eqnarray}
\label{supeq2}
 r=r_h+\frac{r_h^2 N_1}{4}\frac{\rho^2}{c_0^2-z^2} \ , \qquad \cos \theta=\frac{z}{c_0} \ .
\end{eqnarray}

Then 
the standard near horizon behaviour
(\ref{nh1})
of
a generic solution  
translates into 
a well-defined
Schwarzschild-like
rod-structure in Weyl-like coordinates.
For example, 
for $-c_0\leqslant z\leqslant c_0$
and $\rho \to 0$  
one finds the standard timelike rod behaviour discussed above, 
with the leading coefficients in \eqref{rods}-(\ref{nrod2}) given by:
 \begin{eqnarray}
\label{supeq3}
f_{10}(z)=\frac{r_h^2}{c_0^2-z^2} \ , \qquad f_{20}(z)=r_h^2\left(1-\frac{z^2}{c_0^2}\right)\  , \qquad 
f_{02}(z)=\frac{r^{-2\delta_0}N_1^2 r_h^2}{4(c_0^2-z^2)}  \ .
\end{eqnarray}
One can easily verify that the above expressions
imply the same form 
of the Hawking temperature and horizon area
$T_H=e^{-\delta_0}N_1/(4\pi)$
and
$A_H=4\pi r_h^2$,
as found for a Schwarzschild-like line element.
Also, outside this $z-$interval,
$g_{\varphi \varphi}\equiv f_2 \to \rho^2$
 as $\rho \to 0$, while other functions are strictly positive.

\medskip

Further progress can be achieved 
within a numerical approach,
$i.e.$
by computing the same solutions 
in Weyl-type coordinates (\ref{metric2}) and
in the Schwarzschild-like coordinate system
(\ref{Schw-like})
(a case which is technically much simpler,
since  it results in ordinary differential equations rather than PDEs)
and comparing the results.

We have considered this task for 
the case of  spherically symmetric
BH solutions of the considered model (\ref{action})
and an  
{\it even}-parity (spherically symmetric)
 scalar field.
When using the coordinate system (\ref{metric2}),
the problem was solved by using the formalism in 
Section  \ref{2BH},
in particular the ansatz (\ref{ans1})
with a background corresponding to 
the Schwarzschild solution~\eqref{schwey1} and~\eqref{schwey2}. 
Comparing the results found in two different coordinate systems
makes clear that the proposed framework
can be used
to study non-vacuum BHs in Weyl-type coordinates.

\section{An illustration: two BHs balanced by their scalar hair}
\label{results} 
We shall now apply the formalism developed to a concrete case. We shall construct 2BHs configurations, numerically, balanced by their scalar hair. Since we will be dealing with neutral, static BHs, and minimally coupled scalar fields, this requires the scalar potential to violate the weak energy condition in some spacetime regions~\cite{Herdeiro:2015waa}. In subsection~\ref{Sscalar} we provide details about the scalar field model and in subsection~\ref{balance} the constructed solutions are reported.

\subsection{The scalar field potential and scaling properties}
\label{Sscalar} 

We shall assume a $Q$-ball type scalar field 
potential
which is bounded from below,
\begin{eqnarray}
\label{pot}
U(\Phi)=\mu^2 |\Phi|^2  -\lambda |\Phi|^{4}+\nu  |\Phi|^{6} \ ,
\end{eqnarray}
 where $\mu,~\lambda,~\nu$ are
positive constants.
Differently from the canonical $Q$-ball case~\cite{Coleman:1985ki}, 
however, here the scalar field  has no harmonic time dependence.
Importantly, the potential
 {\it is not} strictly positive, 
with  
\begin{equation}
\lambda^2>4 \mu^2 \nu \ .
\end{equation}

Turning now to scaling symmetries of the problem,
we notice first  that
the equations of motion are invariant under the following
 scaling of the  coordinates
$x^i=(\rho,z)$
together with the parameters of the scalar potential   
(in the relations below, the functions or constants which are not mentioned explicitly remain invariant):
\begin{eqnarray} 
\label{symm1}
  x^i= c \tilde x^i \ , \qquad  
 ~\mu=\frac{1}{c}\tilde \mu \ ,  \qquad 
\lambda=\frac{\tilde \lambda }{c^2} \ , \qquad 
\nu=\frac{\tilde \nu }{c^4} \ , 
\end{eqnarray}
with an arbitrary $c>0$. Some relevant quantities
 scale as
\begin{eqnarray} 
  M= c \bar M \ , \qquad 
	T_H= \frac{1}{c}\bar T_H \ , \qquad 
	A_H=c^2 \bar A_H \ , \qquad 
	L=c \bar L \ .
\end{eqnarray}
The equations are also  invariant under a suitable scaling of the
scalar field together with some coupling constants, 
which do not affect the  coordinates:
\begin{eqnarray}
\label{symm2}
\phi = c \tilde \phi  \ , \qquad 
\lambda =\frac{\tilde \lambda }{c^2} \ , \qquad 
\nu =\frac{\tilde \nu }{c^2} \ , \qquad 
G =\frac{\tilde G }{c^2} \ ,
\end{eqnarray}
while 
$
 M= c^2 \tilde M,
$
with
$
T_H,A_H
$
and 
$L$
unaffected.

\begin{figure}[ht]
\hbox to\linewidth{\hss%
	\resizebox{8cm}{6cm}{\includegraphics{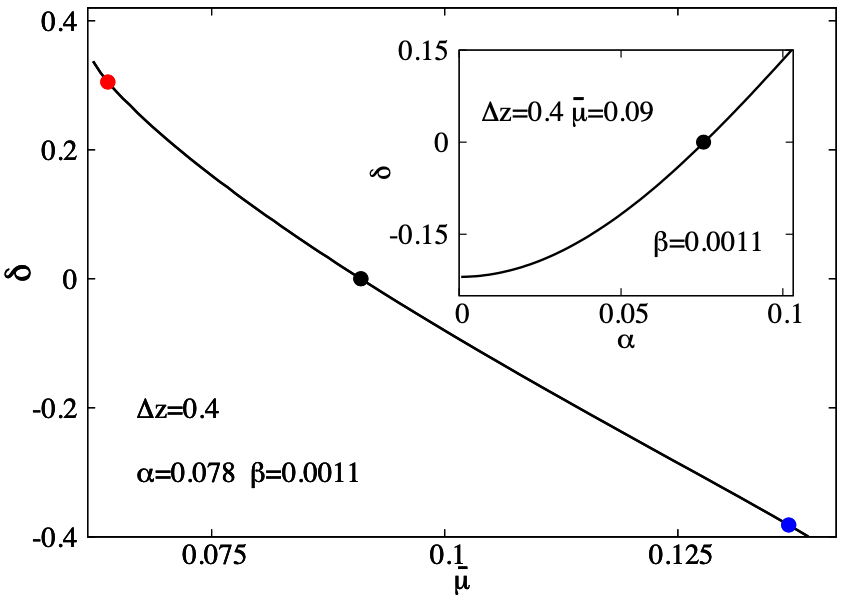}}
\hss}\caption{   
The conical excess/deficit $ \delta$  
as defined by the relation (\ref{delta})  
is shown as a function of the  input parameter
$\bar \mu$ (related to the horizon size)
for 2BHs  with real scalar hair ($m=0$).
The coordinate distance (parameter) between the horizons
$\Delta z$ is fixed, as well the theory parameters $\alpha,\beta$.
The three dots correspond to the solutions displayed in Fig.~\ref{fig2}.
The inset shows a similar plot, with $\delta$ as a function of the theory parameter
$\alpha$
which measures the strength of gravity.
The {\it balanced} configurations are highlighted with black dots.
 }
\label{fig1}
\end{figure}

These symmetries  
are used in practice to simplify the
numerical study of the solutions.
First,
the  symmetry  (\ref{symm1})
is employed to work in units of length set by the scalar field mass,
\begin{eqnarray}
\tilde \mu=1 \ , \qquad i.e. \quad c=\frac{1}{\mu} \ .
\end{eqnarray}
The second symmetry 
(\ref{symm2})
is used 
to set  to  unity the coefficient of the quartic term in the scalar field potential,\footnote{Alternatively,
one can use (\ref{symm2}) to set $\nu=1$ in the potential (\ref{pot}).}
\begin{eqnarray}
\bar \lambda=1 \ , \qquad i.e. \quad c=\frac{1}{\sqrt{\lambda}} \ .
\end{eqnarray}
It follows that two mass scales naturally emerge, 
one set by gravity, $M_{\rm Pl}\equiv 1/\sqrt{G}$,
and the
other one
set by the scalar field parameters, $M_0\equiv \mu{/\sqrt{\lambda}}$.
The ratio of these fundamental mass scales defines the dimensionless coupling constant
\begin{eqnarray}
\label{alpha}
\alpha\equiv \frac{M_0}{M_{\rm Pl}} \ ,
\end{eqnarray}
which is relevant in the physics of the solutions.

Apart from $\alpha$,
the second dimensionless input parameter
is the  (scaled) constant  for the sextic term in the scalar potential, with
 \begin{eqnarray}
\beta\equiv \frac{\nu \mu^2}{\lambda^2} \ .
\end{eqnarray}
Then, the scaled scalar potential reads
$
U(\phi)=\phi^2-\phi^4+\beta \phi^6,
$ 
while 
the Einstein equations
become
$
R_{\alpha\beta}-\frac{1}{2}g_{\alpha\beta}=2\alpha^2 T_{\alpha\beta}.
$

To summarize, after using the scaling symmetries,
the problem still possesses 
four  input parameters:
\begin{eqnarray}
\{ \alpha,~~ \beta\} ~~{\rm and}~~
\{ \Delta z,~~ \bar \mu\} \ , 
\end{eqnarray}
two of them 
determined by the scalar field potential
and the other two by 
 the BW background "seed" solution.
In the solutions with scalar hair,
$\Delta z$ and
$\bar \mu$
are 
still correlated
with, but not strictly corresponding to, the distance between the horizons 
and the horizons mass, respectively.

All quantities shown in this work are given in
natural
 units set by $G$ and $\mu$,
which, in order to simplify the plots, 
we take to unity in what follows.

\begin{figure}[ht]
\hbox to\linewidth{\hss%
	\resizebox{3cm}{6cm}{\includegraphics{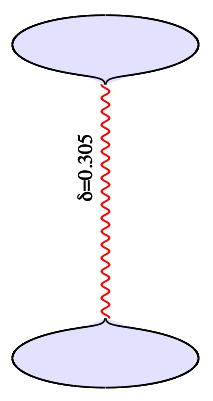}}
\hspace{26mm}%
        \resizebox{3cm}{6cm}{\includegraphics{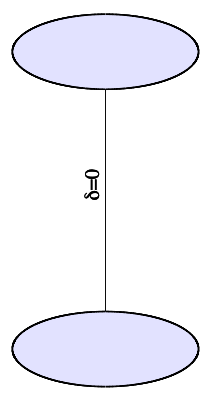}}	
\hspace{26mm}%
        \resizebox{3cm}{6cm}{\includegraphics{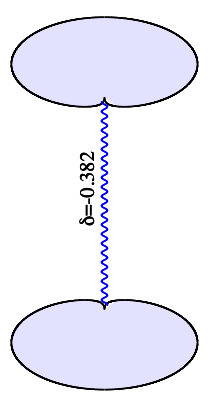}}	
\hss}\caption{  The 
{\it effective}
horizon shape is shown 
for three different 2BH solutions with real scalar hair 
 highlighted with (red, black, and blue) 
dots in Fig.~\ref{fig1}.
The coordinate distance between the horizons is $\Delta z=0.4$,
while 
$\bar \mu=0.064,$
$0.091$,
and
$0.137$,
%
respectively
 (from left to the right).
There is string for $\delta>0$ (leftmost panel) 
and a strut for $\delta<0$ (rightmost panel),
connecting the two horizons. The middle configuration is balanced.
 }
\label{fig2}
\end{figure}
 %
\begin{figure}[ht]
\hbox to\linewidth{\hss%
 \resizebox{8cm}{6cm}{\includegraphics{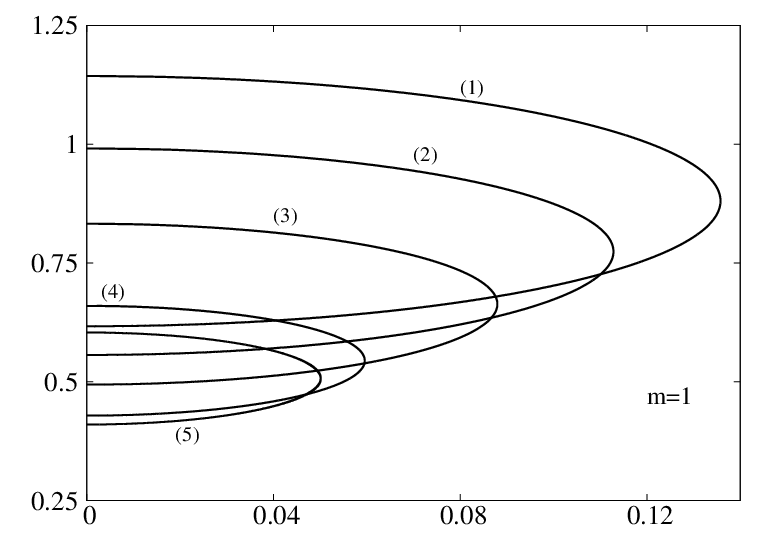}}
\hss}\caption{   
Profiles of the horizon embedding  in
$\mathbb{E}^3$
for the first
("upper") BH
are shown for a set of $m=1$ balanced solutions marked in Fig.~\ref{figure5}. The distance between the BHs decreases from (1) to (5).
}
\label{figSmarr}
\end{figure}

\subsection{The balanced 2BHs system with scalar hair}
\label{balance}

The \textit{decoupling} limit $\alpha \to 0$, where  $\alpha$ defines the 
coupling of the scalar field to gravity as given by (\ref{alpha}), 
corresponds to  
solutions of the scalar field equation
(\ref{eqs2})  on 
a fixed BW geometry,  
$i.e.$~\eqref{metric2} with~\eqref{dictionary} and~\eqref{BW1}-\eqref{BW2} (thus, $ F_i=0$).
Some basic properties of the self-gravitating solutions
are already present in this case.
First, they possess 
 an odd-parity scalar field,  describing a configuration
with two constituents located at
$z=\pm z_0$,
with $z_0$ fixed by the maximum of the energy density,
as resulting from numerics.
For all solutions in this work
(including the ones with scalar back-reaction), we have found 
that $z_0>\Delta z+\bar{\mu}$.
 %
An interesting feature of the decoupling limit analysis is that the scalar 
field never trivializes; 
we could not find any indication 
for the existence of linear scalar clouds on a BW background.
 Also, although more work is necessary,
these {\it non-linear cloud} solutions appear to exist for arbitrary values 
of the BW parameters $\Delta z,\bar{\mu}$.

 \medskip

The backreaction is included by starting
  with the
solutions in the decoupling limit, 
$i.e.$
with given 
scalar field parameter
$\beta$  
and BW background parameters
$(\Delta z, \bar \mu)$, 
and  slowly increasing
 the coupling 
$\alpha$.
Most of the qualitative features of the BW solution  
are still preserved by their  
generalizations with scalar hair obtained in this way.
The generic solutions  possess a conical
singularity which prevents their collapse, and no other pathologies. 
Moreover, by using the formalism 
in \cite{Herdeiro:2009vd}, one can show that
the hairy solutions with conical singularities still admit a
consistent thermodynamical description. 
In particular, when working with the appropriate set of
thermodynamical variables, the Bekenstein-Hawking law still holds,
with the entropy $S=A_H/4$.

The key aspect we wish to emphasise here is the evolution of the conical singularity strength $\delta$ with increasing $\alpha$.
While $\delta<0$ in the decoupling limit,
 one finds that  
$|\delta | $ decreases as $\alpha$ increases,
with the existence of a critical value 
$\alpha_c$ where
$\delta=0$,
$i.e.$ a  balanced configuration.
Moreover, $\delta$ becomes positive 
for $\alpha>\alpha_c$. This behaviour is shown for illustrative values of $(\Delta z,\bar{\mu})$ in the inset of Fig.~\ref{fig1}.

A more physical scanning of the solutions, on the other hand, fixes the theory, $i.e.$ fixes 
($\alpha$,
$\beta$),
which are  constants of the model. 
Then 
a sequences of 
balanced solutions 
in a given theory
can be found by fixing the parameter $\Delta z$ 
(related to the coordinate distance between the two BHs)
and varying the input parameter $\bar \mu$ (related to the BH size).  
 As seen in 
the main panel of 
Fig.~\ref{fig1} 
the system becomes balanced
 for a critical value of the parameter $\bar \mu$, only. For larger (smaller)  $\bar \mu$ the BHs are too heavy (light) and there is a conical excess/strut (deficit/string) in between them. 
This is illustrated in Fig.~\ref{fig2}, where we display
the horizon shape\footnote{
The  Smarr-type embedding in $\mathbb{E}^3$  
of a sequence of balanced $m=1$ solutions 
is shown 
in Figure  \ref{figSmarr}.
}, 
as given by the effective-radius
function ${\rm R}$, relation (\ref{R}),
 for the three  solutions highlighted with (red, black and blue) dots in Fig.~\ref{fig1}.  By repeating this procedure
a (continuous) 
set of balanced solutions 
is found by varying
the  input parameter 
$\Delta z$
and 'shooting' for 
the critical
values
of
$\bar \mu$ which give $\delta=0$.\footnote{Alternatively, 
the same procedure can be performed by interchanging the role
of 
$\Delta z$ and $\bar \mu$.
}
To summarize,
by varying the value of $\Delta z$ 
 and by adjusting the
value of 
$\bar \mu$ via a 'shooting' algorithm, 
the full spectrum of  balanced 2BHs 
with given $(\alpha,\beta)$
can be
recovered numerically, 
in principle. 

\begin{figure}[ht]
\hbox to\linewidth{\hss%
	\resizebox{8cm}{6cm}{\includegraphics{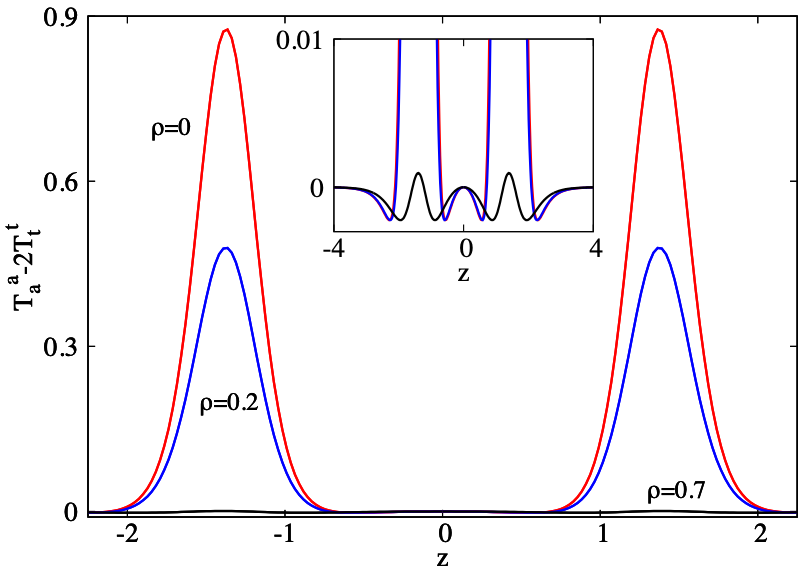}}
\hspace{5mm}%
        \resizebox{8cm}{6cm}{\includegraphics{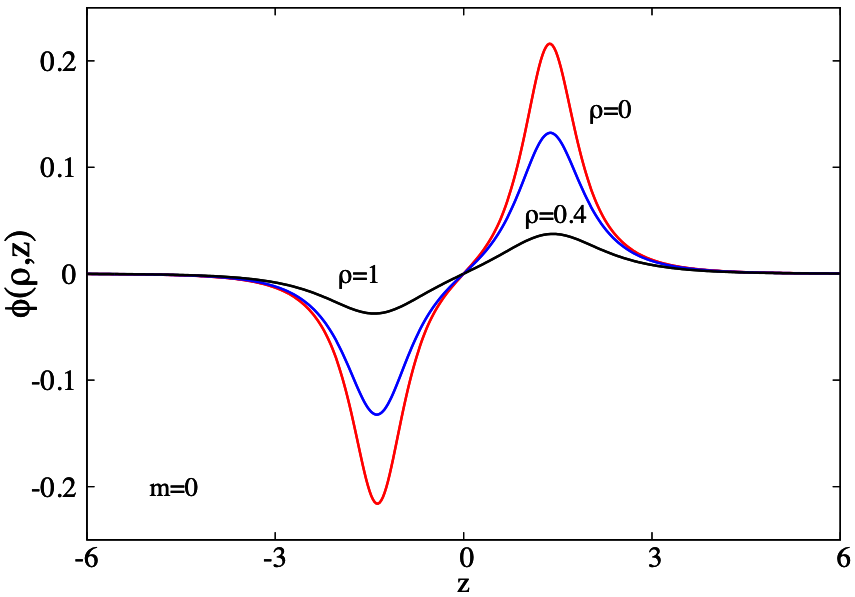}}	
\hss}\caption{   
The  Komar mass-energy density (left panel) and the scalar field amplitude (right panel)   are shown
for a typical balanced  $m=0$ 2BHs system
with scalar hair. The theory constants are 
$\alpha=0.0775$,
$\beta=0.0011$
and geometric input  parameters are
$\Delta z=1.24$,
$\bar \mu=0.071$.
 }
\label{figure4}
\end{figure}

The balanced solutions have no
singularities on and outside the horizon. 
This can be seen by computing the Ricci or the Kretschmann
scalars, which are found to be finite everywhere.
The scalar field is distributed around the two horizons. 
But we have found that the energy
density, as given by the component $ T_t^t$ of the energy-momentum tensor 
always becomes negative in a region around the horizons. This also holds
for 
the Komar mass-energy density
$T_\alpha^\alpha-2T_t^t$ - see the left panel in Fig.~\ref{figure4} --,
although its integral (\ref{Mpsi})
is always positive.
The scalar field profile 
is a also smooth function -- Fig.~\ref{figure4} (right panel).

%
\begin{figure}[ht]
\hbox to\linewidth{\hss%
	\resizebox{8cm}{6cm}{\includegraphics{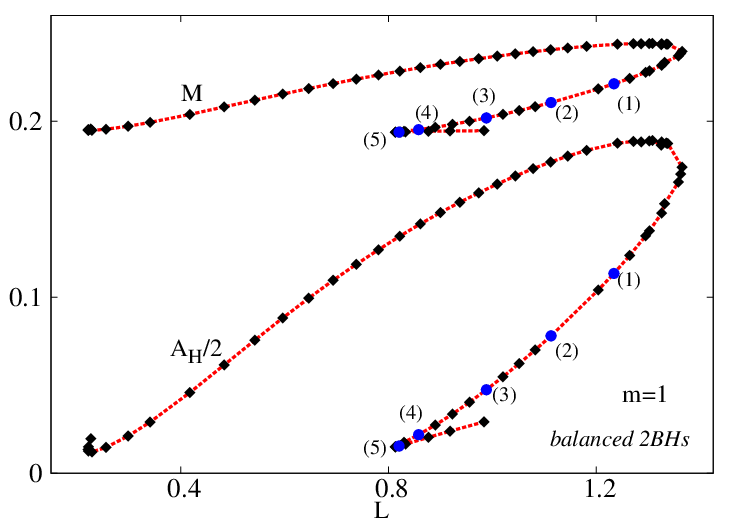}}
\hspace{5mm}%
        \resizebox{8cm}{6cm}{\includegraphics{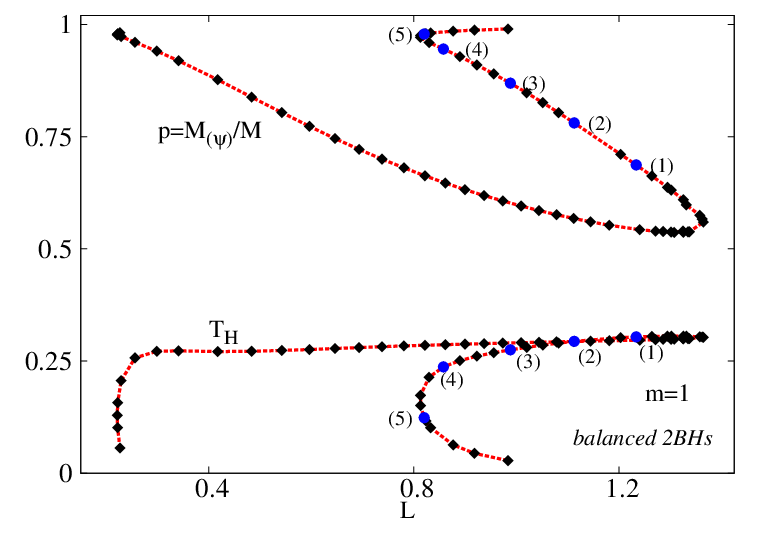}}	
\hss}\caption{  The ADM mass $M$,
the horizon area $A_H$ (left panel),
 the hairiness parameter $p$ 
and the Hawking temperature (right panel)
 of the balanced 
2BHs system with $m=1$, as functions of the
proper distance $L$ between the horizons.
 }
\label{figure5}
\end{figure}

A full scanning of the parameter space of balanced solutions 
is beyond the purposes
of this work. 
We shall focus on balanced solutions
with a fixed set of the theory parameters
$(\alpha=0.0775,$
$\beta=0.0011 )$ 
and $m=1$, which we have studied more systematically\footnote{We have also
constructed a family of balanced BHs
with $m=0$ and the same $(\alpha,\beta)$-values.
However, the limiting behaviour of the solutions is less clear in that case.  }. 
It is reasonable to expect, albeit unproven, 
that one such case captures the basic features of the full space of solutions,
at least for a nonzero $m$.
We emphasise, moreover, that we have established the existence of balanced solutions
for other choices of $(\alpha,\beta)$.

The numerical results indicate that, once the model is fixed, 
there exists a continuous set of  
balanced solutions, which can be labelled by $L$, 
the proper distance between the horizons. 
The most relevant quantities  resulting from numerics 
are shown in 
 Fig.~\ref{figure5}
as a function of $L$
(we recall that all quantities there are given in units set by 
$\mu$ and $G$).
One observes that the solutions exist for a finite range of the proper distance, only
(with $L_{\rm min}>0$).
Thus, although
one may expect to find balanced solutions with an 
arbitrary small or large $L$,
this is not confirmed by numerics.
Moreover, one, two or even three different solution may exist for a 
given proper distance between the horizons,
with a complicated branch structure in terms of $L$.

One observes that all relevant quantities
 vary  significantly as $L$ varies.
Moreover, as expected
from the study in the decoupling limit,
$p$ never vanishes, and it attains a minimal value at $L_{\rm max}$. 
Thus, for any distance $L$, 
a fraction of the spacetime energy must be stored in the scalar hair.
 
Concerning the behaviour of the solutions at the end of the $L$-interval, 
the numerical results suggest the existence of 
two limiting configurations with nonzero values  
for both $M$ and $A_H$,
while the Hawking temperature vanishes, see right panel in  Fig.~\ref{figure5}.
Although we could not  obtain accurate enough solutions 
closer to these critical configurations, 
they are likely singular, 
with a divergent Ricci scalar outside the horizon.

On general grounds, we expect all 2BH balanced 
solutions to be unstable against small perturbation, since their existence requires a fine-tunning between the BHs parameters.
This is supported by the observation that, for any value of the mass,
the configuration maximizing the entropy corresponds to a (single) Schwarzschild  
vacuum BH, and not to a hairy 2BHs solution.

\section{Further remarks}

The asymptotically flat static 2BHs system in ``vacuum" GR necessarily possesses a conical singularity.
A main purpose of this work is to report the existence of static,
balanced  2BHs configurations in Einstein-scalar models, without any electromagnetic fields and keeping asymptotic flatness.
The solutions are supported by the existence of
negative energy densities, 
as allowed by a choice of a scalar potential
which is not strictly positive. 
Our study indicates the existence of
 a one-parameter family of balanced solutions,
 which 
can be parametrized by the physical distance between the horizons.  Our approach is non-perturbative, by solving directly the
Einstein-scalar field equations in a Weyl-type coordinate system,
subject to proper boundary conditions.

In fact, the considered Einstein-scalar theory can be taken as
a simple toy-model for other  cases which may be physically more interesting, for instance avoiding negative scalar energies.
Similar solutions are likely to exist in a variety of other  models,
with similar mechanisms at work.
For example,
effective negative energy densities naturally occur in a 
theories with a Gauss-Bonnet term 
non-minimally
coupled with a scalar field,
see $e.g.$
\cite{Kleihaus:2015aje,Delgado:2020rev}.
Therefore it is natural to conjecture the existence of 
static balanced 2BHs solutions also in such models, although their investigation whould be a more intricate task, by virtue of the complexity of the equations of motion.

Another possible balancing mechanism is to include the effects of rotation. 
This is suggested by the
situation with the black rings in 
 five spacetime dimensions \cite{Emparan:2001wn}.
In that case, the static black ring 
is unbalanced, being supported against collapse 
by conical singularities \cite{Emparan:2001wk}.
Adding rotation balances the ring 
for a critical value
of the event horizon velocity \cite{Emparan:2001wn}.
One may expect that co-rotation of two BHs may help to alleviate the need for negative energies, within
four dimensional (non-vacuum) 
2BH solutions.\footnote{We recall that the double-Kerr ``vacuum" solution
still possess conical singularities, 
see $e.g.$ \cite{Costa:2009wj,Herdeiro:2008kq}
and the references therein.}
Indeed, one may consider  the existence of spinning 
balanced binary BHs in a model 
with a complex massive scalar field 
 {\it without} 
self-interations or negative scalar energies, where the existence of hair is allowed via the synchronization mechanism~\cite{Herdeiro:2014goa,Herdeiro:2014ima}.

 \section*{Acknowledgements}
This work is supported  by the  Center for Research and Development in Mathematics and Applications (CIDMA) through the Portuguese Foundation for Science and Technology (FCT -- Fundac\~ao para a Ci\^encia e a Tecnologia), references  UIDB/04106/2020 and UIDP/04106/2020.  
The authors acknowledge support  from the projects CERN/FIS-PAR/0027/2019, PTDC/FIS-AST/3041/2020,  CERN/FIS-PAR/0024/2021 and 2022.04560.PTDC.  
This work has further been supported by  the  European  Union's  Horizon  2020  research  and  innovation  (RISE) programme H2020-MSCA-RISE-2017 Grant No.~FunFiCO-777740 and by the European Horizon Europe staff exchange (SE) programme HORIZON-MSCA-2021-SE-01 Grant No.~NewFunFiCO-101086251. Computations have been performed at the Argus and Blafis cluster at the U. Aveiro.

\appendix


%

%
\section{Field equations in the parameterization~\eqref{metric2}}
\label{appendix1}

For model~\eqref{action}, with the ansatz~\eqref{metric2} and~\eqref{scalar}, an appropriate combination 
of the Einstein equations,
 $E_t^t=0$,
$E_\rho^\rho+E_z^z=0$
and
$E_{\varphi}^{\varphi}=0$,
yield the following set of equations for the functions 
$f_0,~f_1$ and $f_2$: 
 \begin{eqnarray}
\nonumber
&&
\nabla^2 f_0-\frac{1}{2f_0}(\nabla f_0)^2+\frac{1}{2f_2}(\nabla f_0)\cdot( \nabla f_2)
+ 4 G f_0 f_1 U(\phi)=0 \ ,
\\
\nonumber
&&\nabla^2 f_1-\frac{1}{f_1}(\nabla f_1)^2
-\frac{f_1}{2f_0f_2}(\nabla f_0)\cdot( \nabla f_2)
+4 G f_1  
\left[
(\nabla \phi)^2
-\frac{m^2 f_1 \phi^2}{f_2}
\right]
=0 \ ,
\\
\label{eqs1} 
&&\nabla^2 f_2-\frac{1}{2f_2}(\nabla f_2)^2
+\frac{1}{2f_0}(\nabla f_0)\cdot( \nabla f_2)
 +4 G f_1  (f_2 U(\phi)+2m^2 \phi^2)   =0 \ .
\end{eqnarray}
The equation for the scalar field amplitude $\phi$
is
 \begin{eqnarray}
\label{eqs2n} 
\nabla^2 \phi
+\frac{1}{2f_0}(\nabla f_0)\cdot( \nabla \phi)
+\frac{1}{2f_2}(\nabla f_2)\cdot( \nabla \phi) 
-\frac{m^2 f_1}{f_2}\phi
-\frac{f_1}{2}\frac{dU(\phi)}{d\phi}
=0 \ .
 \end{eqnarray}
%

\begin{figure}[ht]
\hbox to\linewidth{\hss%
 \resizebox{10cm}{6.66cm}{\includegraphics{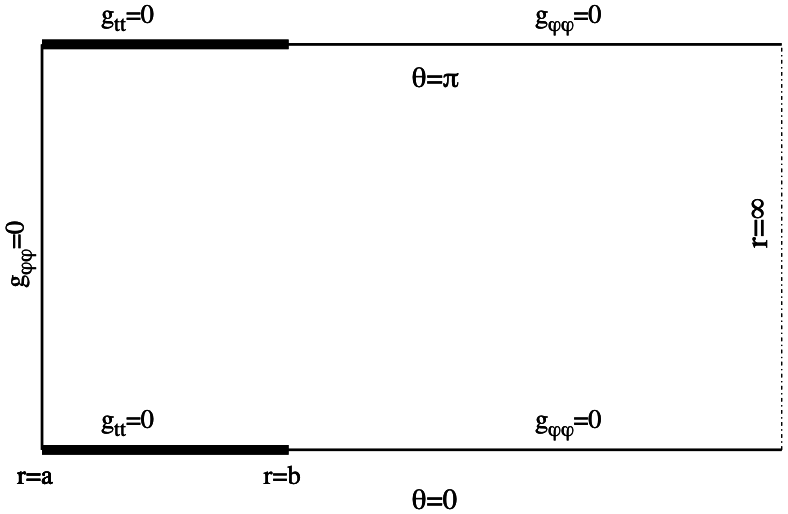}}
\hss}\caption{   
The 2BHs
rod structure in Fig.~\ref{figure1} 
is displayed for $(r,\theta)$-coordinates, 
as defined by Eq. (\ref{coord-transf1}).
}
\label{fig0n}
\end{figure}
%
We have defined, acting on arbitrary functions $\mathcal{F}(\rho,z)$ and  $\mathcal{G}(\rho,z)$, 
\begin{eqnarray}
\label{reln}
(\nabla \mathcal{F}) \cdot (\nabla \mathcal{G}) \equiv \partial_\rho \mathcal{F} \partial_\rho \mathcal{G}+ \partial_z \mathcal{F} \partial_z \mathcal{G} \ , \qquad 
\nabla^2 \mathcal{F}\equiv \partial_\rho^2\mathcal{F}+\partial_z^2 \mathcal{F} \ .
\end{eqnarray}

\section{A new coordinate system
and details on the numerics }
\label{appendix}

Even though the 2BHs solutions reported in this paper can be constructed by employing the 
Weyl-type coordinates $(\rho,z)$, 
the metric ansatz (\ref{metric2}) has a number of disadvantages.
For example,
the coordinate range is unbounded for both $\rho$ and $z$,
which makes it difficult to extract with enough accuracy the value of the mass parameter $M$
from the asymptotic form of
the metric function
 $f_0$.


Thus, in practice, to solve  numerically  the Einstein-scalar field
equations, we have found 
it more convenient to introduce
the new coordinates $(r,\theta)$  
 related to $(\rho,z)$ in (\ref{metric2}) by
\begin{eqnarray}
\label{coord-transf1}
\rho=\frac{r^2-a^2}{r}\sin\theta  \ , \qquad 
z=\frac{r^2+a^2}{r}\cos\theta \ , \qquad {\rm with} \qquad 
\bar \mu=\frac{(a-b)^2}{b} \ ,  \qquad \Delta z=4a \ ,
\end{eqnarray}
with ranges  $a \leqslant r<\infty$
and
$0\leqslant \theta \leqslant \pi$,
and reparametrize the metric (\ref{metric2}) as
\begin{eqnarray}
\label{metric-n}
ds^2=-\hat{f}_0(r,\theta)dt^2+ {\hat{f}_1(r,\theta)}(dr^2+r^2 d\theta^2)
+\hat{f}_2( r,\theta)d\varphi^2 \ .
 \end{eqnarray}

The  rod structure
in  (\ref{metric2}) 
 is still preserved for the new coordinate system, although it becomes less transparent -- 
see Fig.~\ref{fig0n}.  
The two BHs horizons are now located\footnote{
The horizons vanish in the limit $b\to a$,  
in which case 
the coordinate transformation
\begin{eqnarray}
{\cal R}=r \sqrt{1+\frac{a^4}{r^4}+\frac{2a^2 \cos 2\theta}{r^2}},~~
{\cal R}\sin\Theta=\frac{(r^2-a^2)\sin  \theta}{r}.
\end{eqnarray}
results in the  flat spacetime  metic in usual spherical coordinates,
$ds^2 =d{\cal R}^2+{\cal R}^2(d\Theta^2+\sin^2 \Theta d\varphi^2)-dt^2$.
}  at: 
(i) $\theta=0$, for $a\leqslant r\leqslant b$; 
and (ii) 
at
$\theta=\pi$, $a\leqslant r\leqslant b$.
The rod separating the BHs is located at 
 $r=a$ and
$0\leqslant \theta \leqslant \pi$.  
  
Analogous to the case of Weyl-type coordinates,
we define
\begin{eqnarray}
\label{bgf2}
\hat{f}_i=\hat{f}_i^{0} e^{2\hat{F}_i } \ ,
\end{eqnarray}
with the
background functions $\hat{f}_i^{0}$ 
corresponding to the vacuum BW
solution expressed in the $(r,\theta)$-coordinates.
Their explicit expression reads:
 \begin{eqnarray}
  \nonumber
&&
\hat f_0^{0}(r,\theta)= \frac{a^2 b-(a-b)^2r+br^2-2 abr \cos\theta +\sqrt{(b^2+r^2-2br \cos \theta) (a^4+b^2r^2-2a^2br \cos \theta) }}
{-2ab r+a^2(b+r)+br(b+r) -2abr\cos \theta+\sqrt{(b^2+r^2-2br \cos \theta) (a^4+b^2r^2-2a^2br \cos \theta) }}
\\
    \label{f0}
&&{~~}\times
\frac{a^2 b-(a-b)^2r+br^2+2 abr \cos\theta +\sqrt{(b^2+r^2+2br \cos \theta) (a^4+b^2r^2+2a^2br \cos \theta) }}
{-2ab r+a^2(b+r)+br(b+r)+2abr\cos \theta+\sqrt{(b^2+r^2+2br \cos \theta) (a^4+b^2r^2+2a^2br \cos \theta) }} \ ,
 \end{eqnarray} 
\begin{eqnarray}
 \hat f_1^{0}(r,\theta)=\frac{S(r,\theta)\Omega(r,\theta)}{\hat f_0(r,\theta)} \ , \qquad 
\hat f_2^{0}(r,\theta)=\frac{(r^2-a^2)^2\sin^2\theta}{r^2}\frac{1}{\hat f_0(r,\theta)} \ ,
\label{f1}
\end{eqnarray} 
with
 \begin{eqnarray}
&S(r,\theta)=\frac{(a^2-r^2)^2}{2\sqrt{(b^4+r^4-2b^2r^2\cos 2\theta)(a^8+b^4r^4-2a^4 b^2 r^2\cos 2\theta)}}
\\
\nonumber
&\times\frac{a^4b+2a (a^2+b^2)r^2+br^4-(a+b)^2r(a^2+r^2)\cos \theta
+2a^2br^2\cos 2\theta +(a^2+r^2-2ar \cos \theta)\sqrt{(b^2+r^2-2br \cos \theta)(a^4+b^2r^2-2a^2br \cos \theta)}}{a^4+r^4-2a^2r^2\cos 2\theta}
\\
\nonumber
&\times\frac{a^4b+2a (a^2+b^2)r^2+br^4+(a+b)^2r(a^2+r^2)\cos \theta
+2a^2br^2\cos 2\theta +(a^2+r^2+2ar \cos \theta)\sqrt{(b^2+r^2+2br \cos \theta)(a^4+b^2r^2+2a^2br \cos \theta)}}
{a^4b-2a (a^2+b^2)r^2+br^4-(a-b)^2r(a^2+r^2)\cos \theta
+2a^2br^2\cos 2\theta +(a^2+r^2+2ar \cos \theta)\sqrt{(b^2+r^2-2br \cos \theta)(a^4+b^2r^2-2a^2br \cos \theta)}}
\\
\nonumber
&\times\frac{a^4b^2-(a^2+b^2)^2+b^2 r^4
+2a^2b^2r^2\cos 2\theta + \sqrt{(b^4+r^4-2b^2r^2 \cos 2\theta)(a^8+b^4r^4-2a^4 b^2r^2 \cos 2\theta)}}
{a^4b-2a (a^2+b^2)r^2+br^4+(a-b)^2r(a^2+r^2)\cos \theta
+2a^2br^2\cos 2\theta +(a^2+r^2-2ar \cos \theta)\sqrt{(b^2+r^2+2br \cos \theta)(a^4+b^2r^2+2a^2br \cos \theta)}}
\end{eqnarray}
and
\begin{eqnarray}
\Omega(r,\theta)=1+\frac{a^4}{r^4}-\frac{2a^2 \cos 2\theta }{r^2} \ .
\end{eqnarray}

With this parameterization,
we solve  numerically the resulting set of four coupled non-linear
elliptic PDEs for the functions 
$(\hat{F}_i,\phi)$, 
subject to the  set of boundary conditions we now describe.
At $r=a$ one imposes ($i=0,1,2$)
\begin{eqnarray}
\label{ans2}
\partial_r \hat F_i|_{r=a} =0,~\partial_r \phi|_{r=a}=0~~{\rm for}~~m=0,~~{\rm and}~~~
\phi|_{r=a}=0~~{\rm for}~~m\neq 0 \ .
\end{eqnarray}
 The constraint equation ${  E}_{r}^\theta=0$
implies that $\hat F_2-\hat F_1|_{r=a}=$const. 
($i.e.$ a constant value of
the conical deficit/excess
 $\delta$).
At $\theta=0$ the boundary  conditions are
\begin{eqnarray}
\label{ans2z}
\partial_\theta \hat  F_i|_{\theta=0}=            
\partial_\theta \phi |_{\theta=0}=0 \ .
\end{eqnarray}
where again, the constraint equation ${  E}_{r}^\theta=0$
requires  $\hat F_0  -\hat F_1  =$const. for $a\leqslant r\leqslant b$  
($i.e.$ a constant value of the Hawking temperature).
For $r>b$ one finds another supplementary condition, 
$\hat F_2=\hat F_1$, 
thus the absence of a conical
singularity
on the outer $z-$axis,
while for $m\neq 0$
one imposes $\phi |_{\theta=0}=0$ instead of a Newmann boundary condition.
Similar boundary conditions are found for $\theta=\pi.$
At infinity, one imposes the conditions
\begin{eqnarray}
\label{inf}
 \hat F_i|_{r=\infty}= \phi|_{r=\infty}=0 \ ,
\end{eqnarray} 
Moreover, the problem still possesses a $Z_2$-symmetry,
which allows to solve the equations for $0\leqslant \theta\leqslant \pi/2$, only.
The following boundary conditions
are imposed
at $\theta=\pi/2$
\begin{eqnarray}
\label{ans22}
\partial_\theta \hat F_i|_{\theta=\pi/2}=  \phi|_{\theta=\pi/2}=0 \ .
\end{eqnarray}

All numerical calculations  
are performed by using a professional 
finite difference solver,
which uses a  Newton-Raphson method.  
 A detailed presentation of the this code is presented in  \cite{schoen}. 
In our approach, one introduces the new radial variable 
$x = (r-a)/(c+r)$ 
(with $c$ a properly chosen constant)
which maps the semi-infinite
region $[a,\infty)$ to the compact region
$[0,1]$.
The equations for $\hat F_i$ are discretized on a non-equidistant grid in 
$x$
and $\theta$. Typical
grids used have sizes around 
 $200 \times 50$ points, covering the integration region
$[0,1]\times [0,\pi/2]$.
The  numerical error
for the solutions reported in this work is estimated to be typically  $<10^{-3}$. 
However, the errors increase when studying
solutions with small $a$  
and for 
 $b$ 
much larger than $a$
($i.e.$ a large separation of the BHs).

\section{Numerical construction of the  2RNBHs solution }
\label{2RN}

\begin{figure}[ht]
\hbox to\linewidth{\hss%
	\resizebox{8cm}{6cm}{\includegraphics{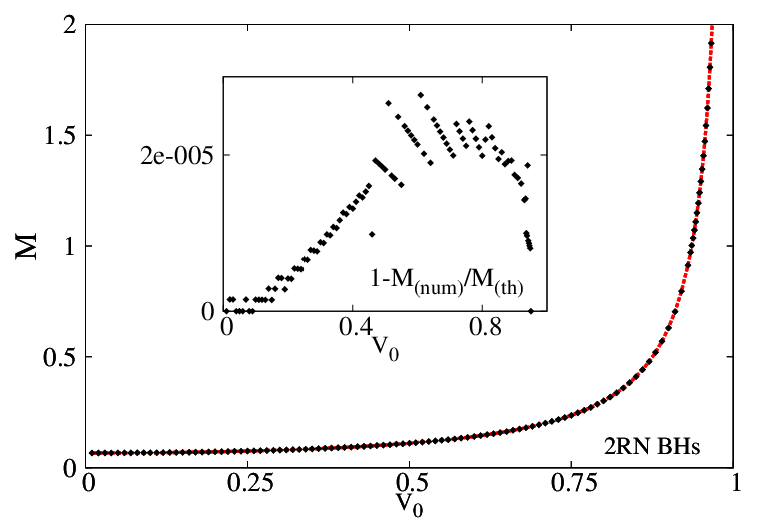}}
\hspace{5mm}%
        \resizebox{8cm}{6cm}{\includegraphics{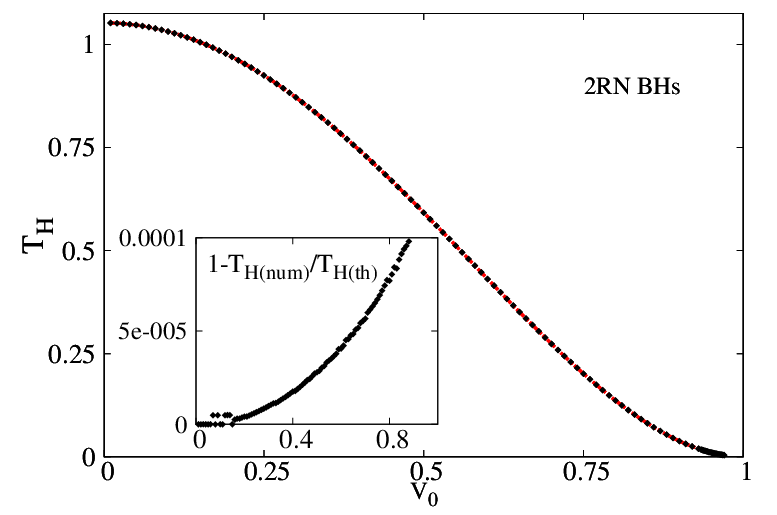}}	
\hss}\caption{  The mass $M$ 
and the Hawking temperature $T_H$
 of the 2RNBHs system 
are shown as a function of
the electrostatic potential $V_0$
for both theory (red line)
and numerical results (dots).
The insets show the relative difference. 
The geometric input parameters are
$a=0.1$,
$b=0.221$.
 }
\label{figRN}
\end{figure}

The simplest application of the proposed formalism consists in recovering 
the double Reissner-Nordstr\"om 
solution by solving numerically  the
Einstein-Maxwell equations
\begin{eqnarray}
R_{\alpha \beta}-\frac{1}{2}g_{\alpha \beta}R=
2 G \left  
(F_{\alpha \gamma}F_\beta^\gamma-\frac{1}{4}g_{\alpha \beta}F^2
\right) \ ,
~~~
\nabla_{\alpha}F^{\alpha \beta}=0 \ ,
\end{eqnarray}
with $F=dA$ the Maxwell field strngth tensor. 

Restricting again to a $\mathbb{Z}_2$-symmetric  solution,
we consider the metric ansatz 
(\ref{metric-n}),   (\ref{bgf2})
and a purely electric Maxwell field,
 \begin{eqnarray}
A=V(r,\theta)dt~,
\end{eqnarray}
and solve numerically the equations for $\hat  F_i$, $V$
by using the approach described above.
The boundary conditions satisfied by the electrostatic potential
$V$ are
\begin{eqnarray}
\label{bcV1}
\partial_r V|_{r=a}=0 ,~~ V|_{r=\infty}=V_0,~~
\partial_\theta  V|_{\theta=\pi/2}=0 ~,
\end{eqnarray}
together with
\begin{eqnarray}
\label{bcV2}
  V|_{\theta=0}=0~~{\rm for}~~a\leqslant r\leqslant b~~{\rm and}
	~~ \partial_\theta V|_{\theta=0}=0~~{\rm for}~~r>b \ .
\end{eqnarray}
The numerical approach is similar to that employed 
for 2BHs with scalar hair,
the input parameters being  the rod coordinates $\{a,b \}$
together with   $V_0$.
However, the corresponding exact solution 
is known in this case\footnote{We recall that the 2RNBHs
solution
can be generated from the 
BW vacuum one by using
a suitable Harrison transformation
(see 
$e.g.$
\cite{Costa:2000kf}).},
and
can be written in the form (\ref{metric-n}),  (\ref{bgf2})
with
\begin{eqnarray}
\label{2RNmetric}
 e^{2 \hat F_0}=\frac{1}{ P^2},~~
 e^{2 \hat F_1}=e^{2 \hat F _1}= P^2,~~
V=\frac{\tanh \gamma}{P}\hat f_0^0,~~
{\rm with}~~P=\cosh^2 \gamma-\sinh^2 \gamma \hat f_0^0 \ ,
\end{eqnarray}
where $\gamma$ is an (arbitrary)  real parameter corresponding to the
electrostatic potential of the configurations,
$\tanh \gamma=V_0$. 
Then 
a straightforward computation leads to the following expression 
of several quantities of interest
\begin{eqnarray}
\label{quant-2RNmetric}
&&
M=\frac{(a-b)^2}{b}\frac{1+V_0^2}{1-V_0^2} \ , \qquad 
 Q_e=\frac{2(a-b)^2}{b}\frac{V_0}{1-V_0^2} \ ,
\\
\nonumber
&&
 A_H=\frac{ 8\pi (a-b)^4(a^2+b^2)}{b^2(a+b)^2}\frac{1}{(1-V_0^2)^2} \ , \qquad 
  T_H=\frac{ b(a+b)^2}{8\pi (a-b)^2 (a^2+b^2)}(1-V_0^2)^2 \ ,~~
\end{eqnarray}
with $Q_e$ the total electric charge\footnote{In this case the value of the U(1) charge is the same for both BHs.}.

In Fig.~\ref{figRN} we show a comparison between 
the theory and numerical results for the mass and Hawking temperature
as a function of  $V_0$
for some fixed $(a,b)$. 
The insets there give an overall estimate for the numerical
accuracy of the solutions, which is consistent with other diagnostics provided
by the solver (similar results hold for $Q_e$ and $A_H$;
 also a similar picture is found when varying $a$ or $b$ at fixed $V_0$).
This supports that the proposed numerical scheme
can be used in the construction of multi-BH solutions
in the presence of matter fields.

 \begin{small}
 
 \end{small}

\end{document}